\documentstyle[12pt]{article}
\title{The relation between the Mandelstam  and the Cayley-Hamilton identities
}
\author {D.E. Berenstein\\
Universidad Nacional de Colombia\\
Departamento de F\'\i{}sica \\
A. A. 14490,  Bogot\'a D.C. \\
Colombia\\
\hbox{\hskip 2cm}\\
and\\
\hbox{\hskip2cm}\\
L.F. Urrutia\\
Instituto de Ciencias Nucleares\\
Universidad Nacional Aut\'onoma de M\'exico\\
Circuito Exterior, C. U.\\
04510 M\'exico, D. F.\\
and\\
Centro de Estudios Cient\'\i{}ficos de Santiago\\
Casilla 16443, Santiago 9
\\Chile}
\begin{document}
\maketitle
\pagebreak
\vskip1pc

\begin{abstract}
\baselineskip=20pt
Starting from the characteristic polynomial for  ordinary matrices we give a
combinatorial deduction of the Mandelstam identities and viceversa, thus
showing that the two
sets of relations are equivalent. We are able to extend  this construction to
supermatrices in such a way that we  obtain the Mandelstam identities in this
case,  once the corresponding  characteristic equation is known.
\vskip 1.5cm
\noindent
PACS: 11.15 \ , \ \  11.30.P \ , \ \  02.
\end{abstract}
\pagebreak
\baselineskip=20pt
\section{Introduction}
When any gauge theory is described in terms of Wilson loops, which are traces
of group elements asssociated to parallel transport around closed space-time
curves (holonomies), one is faced with the problem of having a  non-local and
overcomplete set of variables.  In other words, Wilson loops are constrained
and an important aspect of the kinematics of the problem is just to identify
the reduced phase space  in the loop space of the problem. A most dramatic
example of this situation occurs in 2+1 Chern-Simons theories which are known
to be described by a finite number of true degrees of freedom, which must
arise in this process of reduction from the initially infinite dimensional
phase space.

 An important part of the reduction to the true degrees of freedom is usually
performed by using the Mandelstam identities, whose explicit form depends on
the dimension $n$ of the group matrices and  which provide non-linear
constraints
among the different Wilson loops. These constraints must be solved in order to
exhibit the independent degrees of freedom and this is by no means a simple
problem. The Mandelstam identities can be sistematically derived from the
following identity of $n$-dimensional
$\delta$- functions \cite{Giles},
\begin{equation}
\sum_{P\in{S}_{n+1}}^{}{{(-1)}^{\pi(P)} {\delta}_{{{i}_{1}}{P(j_1)}} \  ... \ \
{\delta}_{{{i}_{n+1}}{P(j_{n+1})}}}=0, \ \ \ \ {i}_{k},{j}_{k}=1, \ ... \  n,
\end{equation}
where the sum runs over all permutations $ P$  of the symmetric group of order
$n+1$ and $ \pi(P)$ denotes the parity of the permutation.  The expression (1)
can be undestood as arising from the expansion,  in terms of a determinant of
$ \delta$-functions,  of the following product of two completely antisymmetric
tensors with $n+1$ indices
$$
\epsilon_{{{i}_{1}}{{i}_{2}} \ ... \ {{i}_{n+1}}}\epsilon _ { {
{j}_{1}}{{j}_{2}} \ ... \ {{j}_{n+1}}},
$$
which nevertheless are allowed to take only $n$ values, thus giving zero as the
result \cite{Gliozzi}.

Contracting $n+1$  holonomy matrices with the relation (1) one obtains a trace
identity among $n+1$ Wilson loops.The resulting Mandelstam identity is
\cite{Giles}
\begin{equation}
\sum_{\hbox{Perm}(1,\dots,n+1)}(-1)^{\pi(P)}W(M_1,\dots,M_{n+1})=0
.\label{mand}
\end{equation}
If the cycle decomposition of the permutation $P$  is given by
$(a(1),\dots,a(i))
\times(a(i+1),\dots)\dots$, then
$$W(M_1,\dots,M_{n+1}) = Tr(M_{a(1)}\dots M_{a(i)}) Tr(M_{a(i+1)}\dots)\dots$$

In relation with the problem of constructing the reduced phase space in
Chern-Simons theories  we can find recently the suggestion that the reduction
process  can be carried over by using   non-linear relations of lower degree
among the traces, derived from the Cayley-Hamilton identity satisfied by the
characteristic polynomial of the group elements involved \cite{Regge,
desitter}. In particular, the reduced phase space
for one genus of a genus $g$ two-dimensional surface in the case of super de
Sitter gravity, which is the Chern-Simons theory of the supergroup $OSp (1|2;
C)$, was obtained   using  a non-linear identity of order four among  the
supertraces of two basic supermatrices.. In this case the supergroup elements
are represented by
$ (2+1)\times(2+1)$ supermatrices and the characteristic polynomial is of
degree 3 \cite{desitter}.

In this work we consider the problem of the equivalence among the two
procedures and we prove that the Cayley-Hamilton identities imply those of
Mandelstam and viceversa in the case of ordinary matrices. We use the same idea
of this proof to give a method for obtaining the Mandelstam identities for
supermatrices, starting from the corresponding characteristic polynomial. These
identities will be important in the reduction of the loop variables phase space
when dealing with gauge theories defined over a supergroup.

The paper is organized as follows: in Section 2 we start from the
characteristic polynomial of an $n \times n$  matrix $M$ and write  explicit
expressions for the coefficients of it in terms of the traces of powers of $M$.
These results are subsequently used in Section 3 to prove  the equivalence
between the Mandeltam and the Cayley-Hamilton identities. Finally, Section 4
contains our construction of the Mandelstam identities in the case of
supermatrices and we present them explicitly  for the simple case of $(1+1)
\times (1+1)$ supermatrices.
\section{Preliminaries}
It is well known that the coefficients of the characteristic polynomial
of an $n \times n$ matrix  $M$ can be written in terms of the traces of the
powers of $M$, up to order $Tr(M^n)$.  If the characteristic equation
of the matrix $M$ is given by
\begin{equation}
P(x)=x^n+a_1x^{n-1}+\dots+a_n
\end{equation}
and if $s_k=r_1^k+\dots+r_n^k$ is the sum of the $k$-th powers of
the roots of $P(x)$, then the Newton equations give a recursive method to
calculate the coefficients $a_i$ \cite{Alge}

\begin{eqnarray}
a_1+s_1&=&0,\nonumber\\
2a_2+a_1s_1+s_2&=&0,\nonumber\\
\vdots \nonumber\\
na_n+a_{n-1}s_1+\dots+s_n&=&0.\label{Newton}
\end{eqnarray}
For any matrix $M$, $s_i$ will be the trace of the $i$-th power of $M$, i.e.
$s_i=Tr(M^i)$.

 An explicit solution of the recursion equations (\ref{Newton}) is given by the
following expression
\begin{equation}
a_i=\sum_{\alpha_1+\dots+\alpha_s=i}{\frac{(-1)^ss_{\alpha_1}\dots
s_{\alpha_s}}{(\alpha_1+\alpha_2+\dots+\alpha_s)(\alpha_2+\dots+
\alpha_s)\dots(\alpha_s)}} \ ,\label{Sol}
\end{equation}
where the sum is made over all the unordered distinct partitions of $i$ and $s$
denotes the total number of terms in  the partition.
The above expression can be demostrated  by using induction, with the
recurrence
\begin{equation}
a_{i+1}=\sum_{k=0}^i{-\frac{a_k s_{i+1-k}}{i+1}}\label{Rec}
\end{equation}
obtained from Eqs.(\ref{Newton}). Substituting  the proposed solution
(\ref{Sol})  for all of the $a_k$, $k<i$,  we obtain
\begin{equation}
a_{i+1}=\sum_{k=0}^i \ \
\sum_{\alpha_1+\dots+\alpha_s=k}{\frac{(-1)^{s+1}s_{\alpha_{i+1-k}}s_{\alpha_1}\dots s_{\alpha_s}}{(i+1) k(k-\alpha_1)\dots(\alpha_s)}} \ . \label{rec1}
\end{equation}
Calling  $ {\alpha}_{0}= \alpha_{i+1-k}$ we realize that the double sum in Eq.
(\ref{rec1}) is just a way of considering all the partitions of $i+1$ into
$s+1$ elements which start with a given $ {\alpha}_{0}$ where $ {\alpha}_{0}
+{\alpha}_{1}\dots+{\alpha}_{s}=i+1$. There are just $i+1$ of them. Besides
$k=i+1-{\alpha}_{0}={\alpha}_{1}\dots+{\alpha}_{s}$ and so on. This leads to
the final result
\begin{equation}
a_{i+1}=\sum_{\alpha_0+\alpha_1\dots+\alpha_s=i+1}{\frac{(-1)^{s+1}s_{\alpha_0}s_{\alpha_1}\dots s_{\alpha_s}}{(\alpha_0+\alpha_1+\dots+\alpha_s)(\alpha_1+\dots+ \alpha_s)\dots(\alpha_s)}} \ .
\end{equation}

Now, it can be shown that expression (\ref{Sol}) can be rewritten in the more
convenient form
\begin{equation}
a_i=\sum_{\begin{array}{c}
k_1\alpha_1+\dots+k_m\alpha_m=i\\
\alpha_1>\alpha_2>\dots>\alpha_m
\end{array}}
\frac{(-1)^{k_1+\dots+k_m}s_{\alpha_1}^{k_1}\dots s_{\alpha_m}^{k_m}}
{k_1!\alpha_1^{k_1}k_2!\alpha_2^{k_2}\dots k_{m}\alpha_s^{k_m}!}\label{prinfor}
\end{equation}
where the sum is  now made over all ordered partitions of $i=\underbrace{
\alpha_1+\dots+\alpha_1}_{k_1}+$ $\dots  +\alpha_m$ and $
{k}_{1}+{k}_{2}+\dots+{k}_{m}$ is the total number of terms in the partition.
In reference \cite{Gliozzi} we can find alternative  expressions for the
coefficients of the characteristic polynomial of an arbitrary matrix  $M$ given
in terms of a recursively defined $n$-entry symbol $ \left\{ {M}_{1},
{M}_{2}\dots{M}_{n}\right\}$, where one sets $
{M}_{1}={M}_{2}=\dots={M}_{n}=M$. The Mandelstam identity for $ n\times n$
matrices is given by $ \left\{ {M}_{1}, {M}_{2}\dots{M}_{n+1}\right\}$=0 in
this formulation.

 As an example of the above expression (\ref{prinfor}) consider the partition
$i=a+b$ in Eq.(\ref{Sol}). Then, permuting $a$ and $b$ so that they become
ordered we can rewrite the  original sum as
$$ \frac{s_as_b}{i\cdot a}+\frac{s_as_b}{i\cdot b}=\frac{s_as_b}{a\cdot b},$$
which corresponds to expression  (\ref{prinfor}) with $ {k}_{1}={k}_{2}=1,
{\alpha}_{1}=a, {\alpha}_{2}=b$.
Nevertheless,  if $a=b$ the two original unordered partitions are the same one,
and we can not count
them twice. The final result in this case is
$$\frac{s_{a}^2}{2a^2},$$
where $ {k}_{1}=2$ and $ {\alpha}_{1}=a$.

We will derive the general result  (\ref{prinfor}) by permuting over the
numbers in a given partition, and then dividing by the  order of the
permutation group that leaves that  partition unchanged.

To begin with we  prove that, for a given unordered partition with $s$ terms,
the sum over the permutations of the
numbers $\alpha_1,\dots,\alpha_s$ in de denominator of Eq.(\ref{Sol}) gives
\begin{eqnarray}
{P}_{s}&=&\sum_{\hbox{Perm$\{\alpha_1,\dots,\alpha_s\}$}}\frac
1{(\alpha_{j(1)}+\alpha_{j(2)}+\dots+\alpha_{j(s)})(\alpha_{j(2)}+\dots+\alpha_{j(s)})\dots(\alpha_{j(s)})}\nonumber\\
&=&\frac1{\alpha_1\dots\alpha_s}.\label{Sumainversos}\end{eqnarray}
Here $j(i)$ is the number in wich $i$ is permuted in the permutation $j$.This
can be shown again by induction calculating the sum over permutations of $ s+1$
elements in the following way. We split the sum into $ s+1$ terms where the
$k-th$ term has $ {\alpha}_{k}$
fixed at the first position while the permutation over the remaining $s$ values
of $ {\alpha}_{s}\neq{\alpha}_{k}$ is calculated according to the expression
(\ref{Sumainversos}). The result of this calculation is
\begin{equation}
{P}_{s+1}=\frac 1{{\alpha}_{1}+{\alpha}_{2}\dots +{\alpha}_{s+1}} \
\frac1{{\alpha}_{2}{\alpha}_{3}\dots{\alpha}_{s+1}}+\frac
1{{\alpha}_{2}+{\alpha}_{1}\dots +{\alpha}_{s+1}} \
\frac1{{\alpha}_{1}{\alpha}_{3}\dots{\alpha}_{s+1}}$$
$$+ \dots
+\frac 1{{\alpha}_{s+1}+{\alpha}_{1}\dots +{\alpha}_{s}} \
\frac1{{\alpha}_{1}{\alpha}_{2}\dots{\alpha}_{s}}.
\end{equation}
Now let us observe that   the denominator containing the sum of  the $
{\alpha}_{k}$'s is common for all the terms.  Besides, ${\alpha}_{k}$ is the
only factor missing in the denominator containing products
in the $k$-th term so that each of these terms can be rewritten as $
{\alpha}_{k}/{{\alpha}_{1}\dots{\alpha}_{k-1}{\alpha}_{k}{\alpha}_{k+1}\dots{\alpha}_{s+1}}$ where  the new denominator is again common for all the terms. Thus the sumation  reduces to those of the
$ \alpha_{k}$'s and the result follows.

 Next we notice that if $\alpha_p$ is repeated
$k_p$ times in the  unordered partition of $i$ into $s$ elements,  the product
$ {\alpha}_{1}{\alpha}_{2}\dots{\alpha}_{s}$ reduces to $
{{\alpha}_{1}}^{{k}_{1}}{{\alpha}_{2}}^{{k}_{2}}\dots{{\alpha}_{m}}^{{k}_{m}}$
where $k_1\alpha_1+\dots+k_m\alpha_m=i$.
Finally the $k_p!$ permutations of these repeated terms produce the same
partition of $i$ so that we must divide by this factor not to overcount  and
the sign is given by the number of terms of the partition, in this case
$s=k_1+\dots+k_s$. It is worthwhile observing that each $a_i$ is an homogeneous
function of $M$ of order $i$.

\section{The relation between the Mandelstam  and the
Cayley-Hamilton identities}
Let us consider the Cayley-Hamilton identity
\begin{equation}
\sum_{i=0}^{n}a_i M^{n-i}=P_M(M)\equiv T_1(M)=0,
\label{pas1}\end{equation}
where we have introduced the notation $P_M$ for the
characteristic polynomial
associated with the matrix $M$, and $a_0=1$. The coefficients $a_i$ have the
explicit
form calculated in Eq.(\ref{prinfor}).  Now,  we can write the following
identity
\begin{equation}
P_{M_1+M_2}({M_1+M_2})=0,
\label{pas2}
\end{equation}
 where the subscript $M_1+M_2$ in $P$ is to emphazise that this substitution is
also performed in the coefficients ${a}_{i}$ of the characteristic polynomial
through Eq.(\ref {prinfor}).
Now we can use Eq.(\ref{pas1}) for $M_1$ and $M_2$ together with
Eq.(\ref{pas2}) to
obtain a reduced identity
\begin{equation}
T_2({M_1,M_2})=P_{M_1+M_2}(M_1+M_2)-P_{M_1}(M_1)-P_{M_2}(M_2)\label{pas3}
\end{equation}

In this  new identity  every term is a homogeneous function
of $M_1$ and $M_2$ of order $n$ and aditionally,  $M_1$ and $M_2$ appear at
least once in every term
of $T_2({M_1,M_2})$. This means that $ T_2({M_1, M_2=0})$ and $
T_2({M_1=0,M_2})$ are identically zero, as can be verified from (\ref{pas3}).
Moreover, we consider that  $T_2({M_1,M_2})$ is fully expanded using the
distributivity of the trace and of the matrix product with respect to matrix
addition.
In a similar fashion, we can construct
\begin{eqnarray}
T_3({M_1,M_2,M_3})&=&P_{M_1+M_2+M_3}(M_1+M_2+M_3)|_{red}\nonumber\\
&=&P_{M_1+M_2+M_3}(M_1+M_2+M_3)
-T_2(M_1,M_2)-T_2(M_1,M_3)\nonumber\\
&&-T_2(M_2,M_3)-T_1(M_1)-T_1(M_2)
-T_1(M_3)=0,\label{pas4}
\end{eqnarray}
which is a sum of null terms.  Again, $ {T}_{3}$ is identically zero when any
of the $ {M}_{i}$'s is set equal to zero. We have introduced the subscript
$|_{{red}}$ to indicate an  identity which has been reduced in such a way that
every matrix involved is present at least once in each term after the identity
is fully expanded.  In other words $ P_{M_1+M_2+M_3}(M_1+M_2+M_3)$ defined in
Eq.(\ref{pas4}) can be directly constructed by expanding the corresponding
characteristic polynomial and discarding all terms in which any one of the
three matrices is missing. Now, let us  define the order $ o(M_i)$ of the
matrix $M_i$ in a monomial of the form

\begin{equation}
aTr(M_1^{\alpha_1}M_2^{\alpha_2}\dots)Tr(M_1^{\beta_1}M_2^{\beta_2}\dots)\dots
M_1^{\gamma_1}M_2^{\gamma_2}\dots
M_1^{\delta_1}M_2^{\delta_2}\dots,\label{order}
\end{equation}
 where some of the exponents may be equal to zero, as the sum of all
of the  exponentes $\alpha_i,\dots$ that appear associated with $M_i$. That is
to say,
$o(M_i)=\alpha_i+\beta_i+\dots+\gamma_i+\delta_i+\dots$.  The
construction of the identity (\ref{pas4}) guarantees that $o(M_i)\geq 1$ for
every
matrix . Is is easy to see that the expanded expression consists of terms like
the one we propose in (\ref{order}).

We can continue in a similar fashion to (\ref{pas4}) and construct
reduced identities of always increasing order, reminding ourselves that
we must substract all  the lower order identities at our dispossal. This
produces a new identity for every order $i=1,\dots, n$.
\begin{equation}
T_k(M_1,\dots,M_k)=P_{M_1+\dots+M_k}(M_1+\dots+M_k)-\sum_{i<k}T_i(M_{s_1},\dots,M_{s_i}),\label{genform}
\end{equation}
where the sum is carried out over all subsets $\{s_1,\dots,s_i\}$ of
$\{1,\dots,k\}$.

Now, $T_{n+1}(M_1,\dots,M_{n+1})$ together with all higher order
identities are identically zero, since
$o(M(i))\geq 1$ for $i=1,\dots,n+1$  would imply that a general term of order
$n+1$ is available in the characteristic polynomial, but we know that the
Cayley-Hamilton identity is only of order $n$.
The identity (\ref{genform}) of order $n$ is very interesting, because every
matrix
must be of order one which means that this particular expression is linear in
each of its components. We will show that $ T_n({M_1,\dots,M_n})$ is
proportional to the Mandelstam identity for $n\times n$ matrices.

The identity formaly reads

\begin{equation}
T_n({M_1,\dots,M_n})=P_{M_1+\dots+M_n}(M_1+\dots+M_n)|_{\hbox{red}}=0,
\end{equation}
which can be  written as
\begin{equation}
\sum_{i=0}^{n}{}a_i(M_1+\dots+M_n)(M_1+\dots+M_n)^{n-i}|_{{red}},
\end{equation}
with $ {a}_{0}=1$.
Now, let us consider in this expression  the contribution
\begin{equation}
\frac{(-1)^{k_1+\dots+k_s}s_{\alpha_1}^{k_1}
s_{\alpha_2}^{k_2}\dots
s_{\alpha_s}^{k_s}}{\alpha_1^{k_1}k_1!\alpha_2^{k_2}k_2!\dots
\alpha_s^{k_s}k_s!}M_{j(1)}M_{j(2)}\dots M_{j(n-i)}\label{red1}
\end{equation}
where the $n-i$ numbers $j(1),\dots,j(n)$ are part of a permutation of the set
$\{1,\dots,n\}$, and naturally $k_1{\alpha}_{1}+\dots+k_s{\alpha}_{s}+n-i=n$.
Now, each of the remaining matrices $M_{j(n-i+1)},\dots,M_{j(n)}$  which are
contained in the corresponding traces,must appear only once in each of the
expanded terms of (\ref{red1}). So, after this reduction the general term will
be of the form

\begin{eqnarray}
&\underbrace{Tr(\overbrace{M_{j(n-i+1)}M_{j(n-i+2)}\dots
M_{j(n-i+\alpha_1)}}^{\hbox{$\alpha_1$ terms}})Tr(M_{j(n-i+\alpha_1+1)}\dots
M_{j(n-i+2\alpha_1)})\dots}_{\hbox{$k_1$-terms}}\nonumber\\
&\dots Tr(M_{j(n-\alpha_s+1)}\dots M_{j(n)})M_{j(1)}
\dots M_{j(n-i)},\label{red2}
\end{eqnarray}
with the coefficient given in (\ref{red1}).

But  we have to add all the terms of the permutations of $\{M_{j(n-i+1)},
\dots,M_{j(n)}\}$ inside the traces
that leave the general term invariant. It is easy to see that this group
of permutations is of order $\alpha_1^{k_1}k_1!\alpha_2^{k_2}k_2!\dots
\alpha_s^{k_s}k_s!$, since it counts the permutations of the
$k_1$ different $s_{\alpha_1}$ where a term can be found $({k}_{1}!)$, and it
also counts the cyclic groups that leave the trace invariant, for example
\begin{equation}
Tr(M_1\dots M_d)=Tr(M_2\dots M_d M_1)
\end{equation}
and all their cyclic permutations $(\alpha_1^{k_1})$.
After this reduction, we  obtain an identity where we are only considering
terms corresponding to inequivalent permutations under the trace and  where all
factors in (\ref{red1}) have cancelled except for the sign. Next we multiply
this espression by ${M}_{n+1} $  and trace it. The  result is

\begin{equation}
\sum_{i}^{n}\sum_{\begin{array}{c}
k_1\alpha_1+\dots+k_s\alpha_s=i\\
\alpha_1>\alpha_2>\dots>\alpha_s
\end{array}} (-1)^{{k}_{1}+\dots+{k}_{s}} \ \  \underbrace{Tr(\overbrace{\dots
\ M \dots
}^{\hbox{$\alpha_1$ terms}})Tr(\dots M\dots)\dots
}_{\hbox{$k_1$ terms}} \ \dots \nonumber $$ $$
\underbrace{Tr(\overbrace{\dots \ M \dots
}^{\hbox{$\alpha_s$ terms}})Tr(\dots M\dots)\dots
}_{\hbox{$k_s$ terms}} \underbrace{Tr(\overbrace{\dots \ M \dots {M}_{n+1}
}^{\hbox{$(n+1-i)$ terms}})}_{\hbox{$1$ term}}=0,
\label{fin1}
\end{equation}
where we emphasize again that the sumation is made only over the inequivalent
cycles of each partition. Now  we rewritte (2\ref{fin1}) in terms of the
partitions of $n+1$ elements into $ {k}_{1}+{k}_{2}+\dots+{k}_{s}+{k}_{s+1}$
cycles with $ {k}_{s+1}=1$ ,  $ {\alpha}_{s+1}=n+1-i$ and where  $
k_1\alpha_1+\dots+k_s\alpha_s+\alpha_{s+1}=n+1$. For fixed $i$, the parity
exponent in (2\ref{fin1}) can be expressed as

\begin{equation}
(-1)^{k_1+\dots+k_s}=(-1)^{n-\pi((k_1\alpha_1)+\dots+(k_s\alpha_s)+(n-i+1))},
\end{equation}
where $\pi((k_1\alpha_1)+\dots+(k_s\alpha_s)+(n+1-i))$ is the parity of the
permutation of $(n+1)$ elements, that corresponds to the  above mentioned cycle
descomposition . This comes about because the parity of each cycle $m$ is $
{\alpha}_{m}+1$ so that the whole parity of the particular descomposition  we
are considering is $ \pi=\sum_{     cycles}^{}
({\alpha}_{m}+1)=n+k_1+\dots+k_s$,  from where the above result follows.
Consider again the cycle descomposition of a permutation of $n+1$ elements in
the form $(k_1\alpha_1)+\dots+(k_s\alpha_s)+(n-i+1)$, and let us further order
the cycles in such a way  that the
$n+1$-th term belongs to the last member of the descomposition. Then we obtain
a term exactly of the form we have previously considered.
Since in every permutation of $n+1$ elements, the $(n+1)$-th term
belongs to some cycle of order $n+1-i$ for some $i$, in practice we are summing
over all the different permutations of $n+1$ elements.
In this way we recover the expression (\ref{mand})  for  the Mandelstam
identities, except
for the sign, that in our case is $(-1)^{n}$ times the one  chosen in Ref.
\cite{Giles}.

Finally,  we  describe how to derive de Cayley-Hamilton identity from the  the
Mandelstam identity. Our starting point is again the expression (2\ref{fin1})
where we set the first $n$ matrices equal to $M$ while considering  $
{M}_{n+1}=X$ to be an arbitrary $n \times n$ matrix . Now we  must count the
number of terms arising from the inequivalent permutations to deduce the
general expression (\ref{prinfor}) for the coefficients of the characteristic
polynomial. Let us focus first on ${a}_{0}$  which arises from the $ i=0$ term
of the sum (2\ref{fin1}). After setting the first $n$ matrices equal to $M$ we
get a factor of $ n!$ for this term. Now let us consider a particular  ordered
cycle descomposition of $i$. The last term in (2\ref{fin1}) contributes with $
\left( {}^{ \  n}_{n-i} \right)(n-i)! $  factors of the type $Tr({M}^{n}X)$.
Now we are left with  $i$ matrices to be distributed among the ${k}_{s}$
remaining cycles. For the $p$ cycle the corresponding factor is
$$
1/{k}_{p}!  \ \left( {}^{ \ {i}_{p}}_{\ {\alpha}_{p}}
\right)({\alpha}_{p}-1)!\left( {}^{ \ {i}_{p}-{\alpha}_{p}}_{ \  \  \
{\alpha}_{p}} \right)({\alpha}_{p}-1)! \dots\left( {}^{ \
{i}_{p}-({k}_{p}-1){\alpha}_{p}}_{ \  \  \  \ \ {\alpha}_{p}}
\right)({\alpha}_{p}-1)! ,
$$
where $ {i}_{p}= i-\sum_{r}^{p-1}{{k}_{r}{\alpha}_{r}}$ is the number of
matrices left out after completing the $ (p-1)$ cycle. Multiplying all these
factors and dividing  the product by  $ {a}_{0}=n!$ we  obtain the result (\ref
{prinfor}) for $ {a}_{i}$. One is lead to the final  expression for the
characteristic polynomial by recalling that
the condition $\sum_{i}^{n}{}{a}_{i}Tr({M}^{n-i}X)=0$ for all
$X$ implies that $\sum_{i}^{n}{}{a}_{i}{M}^{n-i}=0$ . This can be seen by
taking  one by one $X=E_{ij}$,  the standard elements of the basis of the
matrix algebra of $n\times n$ matrices. We have then shown that the Mandelstam
identities and the Cayley-Hamilton identities contain exactly the same
information and  thus they are equivalent.

\section{Mandelstam identities for Supermatrices}

The same construction that we have presented in the case of ordinary matrices
allows for a generalization providing  the Mandelstam identities for the case
of supermatrices.  A definition of the characteristic polynomial  for
supermatrices in terms of $ Sdet(xI-M)$ together with the corresponding
Cayley-Hamilton identity is given in  \cite {Urru}. The latter identity  can be
written in terms of a finite number of supertraces \cite {Jap}. One of  the
differences in this case is that the characteristic polynomial is not monic in
general , so that the coeficient ${a}_{0}$ is a combination of supertraces.
This fact will effectively raise the order of the characteristic polynomial.
In any case, these identities are always homogeneous of some degree, let us say
$t$, in the matrices, and this allows us to make the following definition for
the corresponding Mandelstam identities for supermatrices
\begin{equation}
Str(P_{M_1+\dots+M_t}(M_1+\dots+M_t)|_{{red}}M_{t+1})=0
\label{idstr}.\end{equation}
In this case one also obtains a result that is totally symmetric in the
first $t$ entries, and one would have to prove that it is symmetric in the
$t+1$ entry too. The example we present here posseses complete symmetry in all
the indices
 and we conjecture that it is generally so.  The main drawback
 in the case of supermatrices is the lack of knowledge of a recurrence that
would allow us to obtain a closed expression for the coefficients
of the corresponding characteristic polynomial in terms of supertraces,  in a
manner similar to Eq.(\ref{prinfor}).

As an example, consider the Cayley-Hamilton identity for $(1+1)\times(1+1)$
supermatrices
\cite{Urru}
\begin{equation}
Str(M)M^2-(Str(M^2))M+\frac13(Str(M^3)-Str(M)^3)=0,
\end{equation}
where $t=3$.
 Using this expression  in (\ref{idstr}) we obtain
\begin{eqnarray}
&Str(A)(Str(BCD)+Str(CBD))+Str(B)(Str(ACD)+Str(CAD))\nonumber\\&
+Str(C)(Str(ABD)+Str(BAD))+Str(D)(Str(ABC)+Str(BCA))\nonumber\\&-2Str(AB)Str(CD)-2Str(BC)Str(AD)-2Str(AC)Str(BD)\nonumber\\&
-2Str(A)Str(B)Str(C)Str(D)=0,
\end{eqnarray}
 which corresponds to a symmetric Mandelstam identity of order four . We have
verified this result  using  Mathematica .

The next simple case corresponds to $ (2+1)\times (2+1)$ supermatrices. Here
the characteristic polynomial is of degree $3$ and  order $ t=7$, which will
lead to a Mandelstam identity of order 8 . The final result is not very
illuminating and thus it is not written.

\section{Acknowledgements}

 DEB  would like to acknowledge COLCIENCIAS for their support under
authorization 043-93 that made possible  his stay at  the ICN-UNAM. He also is
indebted to Dr. Roberto Mart\'\i{}nez for constant encouragement.
Both authors  acknowledge support from the project DGAPA-UNAM-100691. LFU is
also supported by the project CONACyT-0758-E9109.

\end{document}